\documentclass[twocolumn]{aastex63}

\usepackage{natbib}
\usepackage{amsmath}
\usepackage{multirow}
\usepackage[normalem]{ulem}
\usepackage{xcolor}
\usepackage{xspace}
\usepackage{enumitem}   
\usepackage{nonfloat}
\usepackage{graphicx}
\usepackage{wrapfig}
\usepackage{lineno}

\newcommand{\Rearth}{$R_\oplus$}

\newcommand{\Msun}{$M_\odot$}

\newcommand{\wotan}{\texttt{W\={o}tan}\xspace}
\newcommand{\epos}{\texttt{epos}\xspace}
\newcommand{\eleanor}{\texttt{eleanor}\xspace}
\newcommand{\gaia}{\emph{Gaia}\xspace}
\newcommand{\pterodactyls}{\texttt{pterodactyls}\xspace}

\newcommand{\degree}{$^{\circ}$}
\newcommand{\kepler}{\emph{Kepler}\xspace}

\newcommand{\ktwo}{\emph{K2}\xspace}

\newcommand{\triceratops}{\texttt{triceratops}\xspace}
\newcommand{\exotic}{\texttt{EXOTIC}\xspace}
\newcommand{\rprs}{$\frac{R_\text{p}}{R_\star}$}

\newcommand{\absg}{$M_\text{G}$}
\newcommand{\extg}{$A_\text{G}$}
\newcommand{\bcg}{$BC_\text{G}$\xspace}
\newcommand{\teff}{$T_\text{eff}$\xspace}

\DeclareUnicodeCharacter{2295}{\oplus}
\DeclareUnicodeCharacter{0301}{\'{e}}

\defcitealias{pecautmamajek2013}{PM13}

\shorttitle{Fernandes, R. B. et al. 2023}
\shortauthors{Fernandes et al. 2023}

\begin{document}
\title{Using Photometrically-Derived Properties of Young Stars to Refine TESS's Transiting Young Planet Survey Completeness}

\correspondingauthor{Rachel B. Fernandes}
\email{rachelfernandes@arizona.edu}

\author[0000-0002-3853-7327]{Rachel B. Fernandes}
\altaffiliation{President's Postdoctoral Fellow}
\affil{Department of Astronomy \& Astrophysics, Center for Exoplanets and Habitable Worlds, The Pennsylvania State University, University Park, PA 16802, USA}
\affil{Lunar and Planetary Laboratory, The University of Arizona, Tucson, AZ 85721, USA}
\affil{Alien Earths Team, NASA Nexus for Exoplanet System Science, USA}

\author[0000-0003-3702-0382]{Kevin K. Hardegree-Ullman}
\affil{Steward Observatory, The University of Arizona, Tucson, AZ 85721, USA}
\affil{Alien Earths Team, NASA Nexus for Exoplanet System Science, USA}

\author[0000-0001-7962-1683]{Ilaria Pascucci}
\affil{Lunar and Planetary Laboratory, The University of Arizona, Tucson, AZ 85721, USA}
\affil{Alien Earths Team, NASA Nexus for Exoplanet System Science, USA}

\author[0000-0003-4500-8850]{Galen J. Bergsten}
\affil{Lunar and Planetary Laboratory, The University of Arizona, Tucson, AZ 85721, USA}
\affil{Alien Earths Team, NASA Nexus for Exoplanet System Science, USA}

\author[0000-0002-1078-9493]{Gijs D. Mulders}
\affil{Facultad de Ingenier\'{i}a y Ciencias, Universidad Adolfo Ib\'{a}\~{n}ez, Av.\ Diagonal las Torres 2640, Pe\~{n}alol\'{e}n, Santiago, Chile}
\affil{Millennium Institute for Astrophysics, Chile}
\affil{Alien Earths Team, NASA Nexus for Exoplanet System Science, USA}

\author[0000-0001-6476-0576]{Katia Cunha}
\affil{Observat\'{o}rio Nacional/MCTIC, R. Gen. Jos\'{e} Cristino, 77, 20921-400, Rio de Janeiro, Brazil}
\affil{Steward Observatory, University of Arizona, 933 North Cherry Avenue, Tucson, AZ 85721-0065, USA}
\affil{Institut d’Astrophysique de Paris, UMR7095 CNRS, Sorbonne Universit\'{e}, 98bis Bd. Arago, 75014 Paris, France}

\author[0000-0003-2008-1488]{Eric E. Mamajek}
\affil{Jet Propulsion Laboratory, California Institute of Technology, 4800 Oak Grove Drive, Pasadena, CA 91109, USA}
\affil{Department of Physics and Astronomy, University of Rochester, Rochester, NY 14627-0171, USA}

\author[0000-0002-5785-9073]{Kyle A. Pearson}
\affil{Jet Propulsion Laboratory, California Institute of Technology, 4800 Oak Grove Drive, Pasadena, CA 91109, USA}

\author[0000-0002-2012-7215]{Gregory A. Feiden}
\affil{Department of Physics and Astronomy, University of North Georgia, Dahlonega, GA 30597 USA}

\author[0000-0002-2792-134X]{Jason L. Curtis}
\affil{Department of Astronomy, Columbia University, 550 West 120th Street, New York, NY 10027, USA}


\begin{abstract}
The demographics of young exoplanets can shed light onto their formation and evolution processes. Exoplanet properties are derived from the properties of their host stars. As such, it is important to accurately characterize the host stars since any systematic biases in their derivation can negatively impact the derivation of planetary properties. Here, we present a uniform catalog of photometrically-derived stellar effective temperatures, luminosities, radii, and masses for 4,865 young ($<$1\,Gyr) stars in 31 nearby clusters and moving groups within 200\,pc. We compared our photometrically-derived properties to a subset of those derived from spectra, and found them to be in good agreement. We also investigated the effect of stellar properties on the detection efficiency of transiting short-period young planets with TESS as calculated in \citet{fernandes2022}, and found an overall increase in the detection efficiency when the new photometrically derived properties were taken into account. Most notably, there is a $1.5\times$ increase in the detection efficiencies for sub-Neptunes/Neptunes (1.8--6\,\Rearth) implying that, for our sample of young stars, better characterization of host star properties can lead to the recovery of more small transiting planets. Our homogeneously derived catalog of updated stellar properties, along with a larger unbiased stellar sample and more detections of young planets, will be a crucial input to the accurate estimation of the occurrence rates of young short-period planets.

\end{abstract}


\section{Introduction} \label{sec:intro}
The discovery of thousands of transiting exoplanets via large-scale surveys such as \kepler \citep{Borucki2010}, \ktwo \citep{howell2014k2}, and TESS \citep{ricker2014transiting} has enabled us to explore not only individual planets and systems, but also planet populations. However, the determination of exoplanet properties are dependent on robust knowledge of their host stars, most notably stellar radii, masses, and effective temperatures which can be used to derive planet radii, masses, and equilibrium temperatures. In order to facilitate target selection, catalogs providing basic stellar properties were developed for \kepler, \ktwo, and TESS such as the \kepler Input Catalog \citep[KIC,][]{Brown2011}, the Ecliptic Plane Input Catalog \citep[EPIC,][]{Huber2016}, and the TESS Input Catalog \citep[TIC,][]{Stassun2018}, respectively. These input catalogs were derived by combining data from heterogeneous sources, and hence cannot be used as a precise reference for stellar properties and exoplanet demographics.

For studies of exoplanet populations, it is especially crucial to have a homogeneously derived stellar catalog as any systematic biases in the derivation of the stellar properties can negatively impact the derivation of planetary properties, and therefore lead to the incorrect characterization of the planet populations. One prominent example is the discovery of the radius valley in \kepler's short-period small planet population, which was enabled by uniform spectroscopic \citep{fulton2017california} and astroseismic \citep{van2018asteroseismic} stellar classification. The \gaia mission \citep{GaiaCollaboration2016} has transformed the field of stellar classification, providing photometry and distances to over a billion stars \citep{GaiaCollaboration2018,GaiaCollaboration2021}. With \gaia DR2 data, \citet{Berger2020} and \citet{hardegree2020scaling} were able to update the \kepler and \ktwo stellar catalogs, respectively, in a homogeneous manner, with the latter enabling the confirmation of the planet radius valley beyond the \kepler field. \citet{Stassun2019} also incorporated \gaia DR2 data into the TIC. However, there are still many inhomogeneities in the derivation of TIC stellar properties, so it should not be used for large-scale exoplanet demographic studies. A fully uniform catalog of TESS stellar properties for nearly two billion stars is a difficult task, but uniform catalogs of subsets of TESS targets are a much more tractable goal.

While most known short-period transiting exoplanets have been found orbiting Gyr-old stars, over the past decade \ktwo and TESS have facilitated the discovery of $>$30 young ($<$1\,Gyr)  exoplanets \citep[A. Vanderburg priv. comm.; e.g.,][]{newton2019tess,newton2021tess, rizzuto2020tess,mann2020tess,nardiello2020psf,bouma2020cluster}. The discovery and characterization of young short-period exoplanets is crucial to our understanding of planet formation and evolution processes such as photoevaporation \citep[e.g.,][]{owen2013kepler, owen2017evaporation}, and core-powered mass loss \citep[e.g.,][]{ginzburg2016super, ginzburg2018core,gupta2019sculpting, gupta2020corecool, gupta2020signatures}. Young stellar clusters and moving groups are promising targets to search for such planets. TESS enables this search by providing light curves spanning nearly the entire sky. However, if we hope to learn about the demographics of young planets, we still need a uniform catalog of stellar properties. Young stars are typically still evolving onto the main sequence, so their properties cannot necessarily be derived using the same assumptions, inputs, or relationships as their main sequence counterparts. While past studies of individual young clusters and associations have yielded stellar properties for those clusters \citep[e.g.,][]{Fang2023}, we aim to study planets in all known nearby young clusters and associations (within 200\,pc), and this necessitates a larger homogeneous catalog.

Ideally, we would have a spectrum for each star to yield precise spectral type, effective temperature, surface gravity, and metallicity information. However, less than 10\% of our sample of young stars have spectra and so we must depend on available photometry to characterize these stars. Here, we present a uniformly derived catalog of photometrically-derived stellar properties for 4,865 stars in nearby (within 200\,pc) young clusters and moving groups. In Section~\ref{sec:classify}, we used data from \gaia DR3 to derive stellar effective temperatures, luminosities, radii, and masses. We tested our derived stellar properties by comparing them to properties measured from spectra for a subset of our sample in Section~\ref{sec:compare}. Next, we used our stellar catalog to update the detection efficiency analysis and planet occurrence rate calculations for young stars from \citet{fernandes2022} in Section~\ref{sec:completeness}. Finally, we summarized our results and discuss the next steps needed to further improve our stellar catalog and planet occurrence rates for young stars in Section~\ref{sec:summary}.

\section{Stellar Classification}\label{sec:classify}

The focus of this work is to compute a homogeneous catalog of stellar \teff, radii, and masses to aid in the accurate estimation of young transiting exoplanet radii as well as to place constraints on their demographics. Given that TESS decreases in sensitivity to transiting planets orbiting fainter low-mass stars \citep[e.g.,][]{Dietrich2023}, we only classified stars of F, G, K, and early M spectral types (down to M3.5~V) as they are typically bright enough and have a higher chance of hosting a detectable planet with TESS.

Our sample of nearby moving groups and young clusters was compiled using the BANYAN $\Sigma$ \citep{gagne2018banyan}, and the \gaia DR2 open cluster member lists (\gaia Collaboration, \citealt{babusiaux2018gaia}). We also added the more recently discovered Argus \citep{Zuckerman2019}, MUTA \citep{gagne2020}, and Pisces-Eridanus \citep{Curtis2019} groups. We restricted the median moving group or young cluster distance to $\sim$200\,pc to ensure that we can detect planets around later K- and early M-type stars with TESS. We excluded clusters younger than 10\,Myr since their stars could still retain a disk \citep[e.g.,][]{ErcolanoPascucci2017}, and their light curves are highly complex and variable \citep[e.g.,][]{Cody2014}. We included clusters up to 1\,Gyr to cover ages over which the short-period planet population is expected to evolve \citep[e.g.,][]{Rogers2021}. With these distance and age cuts, we obtained a starting sample of 10,585 young stars from 31 young clusters and moving groups (see Table~\ref{table:photometry} in Appendix~\ref{sec:photometry}). 

\subsection{Main Sequence vs. Pre-Main Sequence Sample}\label{sec:msvspms}

The age at which any given star reaches the main sequence depends on its stellar mass: a 1.6\,\Msun\ F-type star takes $\sim$20\,Myr to reach the main sequence, whereas a 0.1\,\Msun\ M dwarf (like TRAPPIST-1) can take up to one billion years. However, the majority of our sample of young stars do not have measured masses, which makes it challenging to determine which stars in a given cluster are pre-main sequence and which stars have already reached the main sequence. To combat this, we relied on $G$, $G_{BP}$, and $G_{RP}$ band magnitudes for our targets from \gaia\ DR3 \citep{gaiadr3}, along with distances from \citet{bailerjones2021}, and 2MASS IDs and photometry \citep{Skrutskie2006} from \gaia's best neighbour cross-match (see Table~\ref{table:photometry} in Appendix~\ref{sec:photometry}). At this stage, we removed all targets with non-finite magnitudes, parallaxes, and associated errors since we could not compute stellar properties for them, leaving us with 10,238 targets.

In crowded fields (typical in young cluster environments), it is highly probable that a given exoplanet transit is diluted due to the light from a bound companion star which can lead to an underestimated planet radius measurement. To this effect, we identified and removed any known binaries and non-single sources in our sample using the \gaia EDR3 binary catalog \citep{El-Badry2021}, the Robo-AO census of companions within 25\,pc \citep{Salama2022}, and the \gaia\ DR3 non-single stars catalog \citep{GaiaCollaboration2022binaries}, leaving us with 9,462 targets. \gaia also provides a Renormalised Unit Weight Error (RUWE) score for each source. For sources where the astrometric observations fit well with the single-star model, the expected RUWE value is around 1.0. However, a value greater than 1.0 suggests the source could be either non-single or has other issues that may affect the accuracy of the astrometric solution. Targets with a RUWE score $>1.4$ have been found to be indicative of a non-single source \citep{Ziegler2020}. As such, we also removed any stars with a \gaia DR3 RUWE score $>1.4$, which left us with 8,239 stars for which stellar properties could be computed. 

Here, it is important to note that although we removed all known binaries and sources with high RUWE scores from our sample, there is still a possibility of unresolved binaries leading to improper star classification. For instance, if a G-type main sequence star has an unresolved equal-mass binary, the photometric colors would not be affected, but the combined flux from both stars would lead to an overestimation of the system's luminosity. This overestimation of the luminosity would propagate into the mass and radius estimates derived from the luminosity, leading to an overestimation of both quantities. Since pre-main sequence stars are more luminous and have larger radii, the mischaracterization of a main sequence star as pre-main sequence could occur if there is an unresolved binary. This mischaraterization would further propagate into the derivation of the planet radii. More specifically, an equal mass binary would cause the luminosity to be overestimated by 2$\times$. This means that the radius of the star as derived from the flux would be $\sqrt{2}\times$ larger than the true radius. This would cause the transit depth of the planet to be diluted by 0.5$\times$, and the radius of the planet to be underestimated by $\sqrt{2}\times$ or approximately 41.42\%. Assuming a stellar multiplicity rate of 44\% and 26\% for FGK and M stars respectively \citep{duchene2013stellar}, and given that our previous known binary and high RUWE score cuts removed $\sim$20\% of the sample, we are left with a possible $\sim$292 to 1,168 unresolved binaries in our sample that are likely mischaracterized. Dilution resulting from unresolved binaries is also expected to hinder our ability to detect the transit signals of smaller planets, ultimately reducing our detection efficiency within those particular bins. To address this issue, performing a thorough assessment of the multiplicity of each star in our sample would be ideal. However, it would necessitate extensive ground-based follow-up using high-resolution imaging, which is beyond the scope of this paper.

\begin{figure}[!htb]
    \centering
    \includegraphics[width=0.95\linewidth]{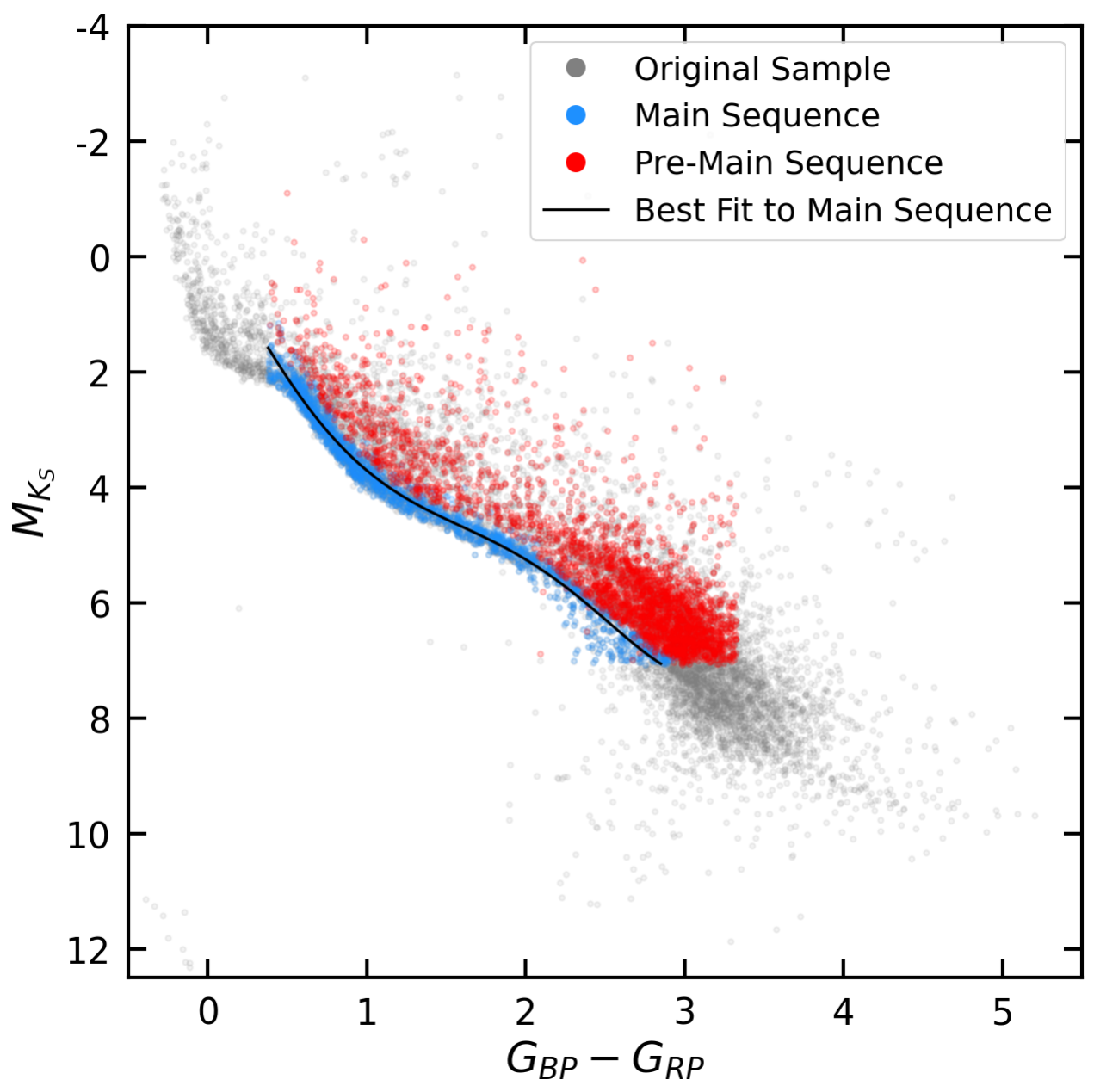}
    \caption{Hertzsprung-Russell diagram of our starting sample of stars from 31 young clusters and moving groups (grey). Stars classified in this work as pre-main sequence and main sequence stars are also shown in red and blue, respectively. Around $M_{K_S}$=2\,mag there is an apparent bifurcation in the main sequence stars. The sparsity of data points in that region for which to fit our color-magnitude polynomial could lead to some misclassification of pre-main sequence stars as main sequence stars, which we think is happening here.}\label{fig:hrdiagram}
\end{figure}

Using the intrinsic color of a star i.e., the difference in magnitude calculated using two different color filters, along with the absolute magnitude, we created a Hertzsprung-Russell diagram \citep{Hertzsprung1911,Russell1914} to differentiate between the pre-main sequence and main sequence stars (see Figure~\ref{fig:hrdiagram}). 

Here, we used \gaia DR3 and 2MASS photometry of the standard main-sequence stars that were used to create Table 5\footnote{An updated version is maintained at this site: \url{https://www.pas.rochester.edu/\~emamajek/EEM\_dwarf\_UBVIJHK\_colors\_Teff.txt}} from \citet[][henceforth PM13]{pecautmamajek2013} in order to establish $M_{K_S}$ vs. $G_{BP}-G_{RP}$ or $G-G_{RP}$ relationships. We determined the main sequence by fitting a 5$^{\mathrm{th}}$ order polynomial to $M_{K_S}$ and $G_{BP}-G_{RP}$. The polynomial order for each of the fits in this paper was the lowest order needed to fit large-scale structure while higher orders did not significantly improve the RMS residual scatter. Since $G_{BP}$ is not well constrained for intrinsically redder stars, we instead fit $M_{K_S}$ and $G-G_{RP}$ with a 5$^{\mathrm{th}}$ order polynomial for targets with $4.5 < M_{K_S}< 10$. All stars within 10\% of both these fits were analyzed as main-sequence stars in this work, while the rest were analyzed as pre-main sequence i.e. stars of a given color are labeled as main sequence if their magnitude is within 10\% of that found by evaluating the main sequence best-fit polynomial at that color. At this point, we also removed stellar populations that we could not accurately classify for both main sequence and pre-main sequence populations\footnote{Table 6 of \citetalias{pecautmamajek2013} is only complete to M5 pre-main sequence stars, which we used as a guide for our color and magnitude limits.}: white dwarfs and OBA stars ($0.5 <G_{BP}-G_{RP}<3.5$), and late M stars ($M_{K_S}>7.1$). These cuts gave us a total population of 4,865 stars (1,824 main sequence, and 3,041 pre-main sequence) for which we derived stellar properties (see Figure~\ref{fig:hrdiagram}).


\subsection{Stellar Effective Temperatures and Radii}\label{sec:radii}
Using intrinsic colors, we can derive stellar effective temperature (hereafter, \teff). The \teff for pre-main sequence stars was computed by fitting a 7$^{\mathrm{th}}$ degree polynomial to the derived $G_{BP}-G_{RP}$ colors and \teff for pre-main sequence stars from \citetalias{pecautmamajek2013} Table~6. We note that the \citetalias{pecautmamajek2013} pre-main sequence table does not have \gaia\ color information, but it has $V-I_{c}$ colors, which we converted to $G_{BP}-G_{RP}$ using \gaia photometric relations with other photometric catalogs \citep{gaia_edr3_ch5_2021}. For main sequence stars, we fit a 4$^{\mathrm{th}}$ order polynomial to $G_{BP}-G_{RP}$ versus \teff from the set of individual measurements from standard stars used to create the most recent version of the \citetalias{pecautmamajek2013} Table~5.

All uncertainties herein were calculated using a Monte Carlo method. For \teff, we started with each $G_{BP}$ and $G_{RP}$ magnitude and associated error, drew a normal distribution of 1000 points, computed $G_{BP}-G_{RP}$ from those posteriors, ran the $G_{BP}-G_{RP}$ posterior through the polynomial fit, and took the median value of the resultant distribution as our \teff. We added the standard deviation of the \teff distribution in quadrature to the $\sim$1\% and $\sim$1.66\% rms deviation from the polynomial fits for pre-main main sequence stars, respectively. The median uncertainties on \teff are 1.34\% and 2.67\% for pre-main and main sequence stars, respectively. The main sequence \teff uncertainties are larger because there were significantly more data points used for that polynomial fit, resulting in a larger rms scatter. Our \teff uncertainties are likely underestimated, but they are the best we can do given the data we have for our polynomial fits. We urge caution using these \teff uncertainties for individual targets.

In Section~\ref{sec:msvspms}, we used $M_{K_S}$ magnitudes to define our sample of main and pre-main sequence stars, particularly to help establish cutoffs for late M dwarfs. Since we are classifying stars from F to early M, we chose to apply bolometric corrections in the optical $G$ band to minimize these corrections across different spectral types. We calculated the bolometric magnitude using the absolute $G$ magnitude (\absg), the bolometric correction in $G$ band (\bcg), and the total extinction in $G$ band (\extg) as in the equation below:

\begin{equation}
    M_\text{bol} = M_\text{G} + BC_\text{G} - A_\text{G}
\end{equation}

\noindent Since we did not immediately have a value for $BC_\text{G}$, we converted \citetalias{pecautmamajek2013}'s $BC_\text{J}$ to \bcg. We first fit a 7$^{\mathrm{th}}$ degree polynomial to the $G_{BP}-G_{RP}$ colors we computed earlier and $BC_\text{J}$ from \citetalias{pecautmamajek2013} Table~6 and used that relation to compute $BC_\text{J}$ for our targets. Then we converted $BC_\text{J}$ to \bcg with \bcg $= BC_\text{J} - (m_G - m_J)$, where $m_G$ and $m_J$ are the apparent magnitudes in \gaia G and 2MASS J bands, respectively. 
We then computed the bolometric luminosity as follows:
\begin{equation}
    L_\text{bol} = L_\text{0} \times 10^{-0.4 \times M_\text{bol}}
\end{equation}
\noindent where $L_{0} = 3.0128 \times 10^{28}$ W is an IAU value also used in \citet{mamajek2015}. We accounted for extinction in each band by computing reddening using the \texttt{dustmaps} code \citep{Green2018}. For targets north of $-30$\degree\ in declination, we used the most recent Bayestar map \citep{Green2019} to compute reddening in $J$-band and converted it to extinction by multiplying the value by 0.7927 from Table~1 of \citet{Green2019}. For targets south of $-30$\degree\ in declination, which the Bayestar map does not cover since it was calibrated using data from the northern hemisphere Pan-STARRS survey, we used the recalibrated \citet{Schlegel1998} dust map from \citet{Schlafly2011}. For these targets we computed $E(B-V)$ using \texttt{dustmaps} and converted it to $A_V$ extinction by multiplying the value by 2.742 (assuming $R_V=3.1$) from Table~6 of \citet{Schlafly2011}. We then converted $A_J$ and $A_V$ to extinction in other bands by multiplying our values by those listed in Table~3 of \citet{wangchen2019}, again assuming $R_V=3.1$. Since our targets are typically within 200\,pc, the effect of extinction is minimal (median $A_V=0.03$\,mag, $A_J=0.007$\,mag), so we are not concerned about using two different dust maps for northern and southern targets, or converting extinctions to different bands.

Using our \teff and luminosity values, we then computed stellar radius using the Stefan-Boltzmann law:
\begin{equation}
R_{\star} = \sqrt[\leftroot{-2}\uproot{5}]{\frac{L_\text{bol}}{4\pi\sigma_\text{sb}T_{\text{eff}}^4}}
\end{equation}
\noindent where $\sigma_\text{sb}$ is the Stefan-Boltzmann constant. However, for lower mass or ``redder'' targets in the range $4.5 < M_{K_S} < 7.1$, the stellar radii were computed using the magnitude--radius relationship from \citet{Mann2015}. This magnitude--radius relationship was calibrated using 183 well-characterized nearby K7--M7 single stars and yields radius uncertainties of $\sim$3\% for the above $M_{K_S}$ magnitude range. FGK stars have higher median radius uncertainties of 3.7\% and 5.7\% for pre-main and main sequence stars, respectively, but we again urge caution with these uncertainties since they depend partially on our low \teff uncertainties.


\subsection{Stellar Masses}
In order to derive stellar masses for the main sequence stars in our sample, we used the mass--luminosity relation from \citet{Torres2010} for stars with $\log(L_{\star}/L_{\odot})>-1$, which corresponds to a mass $M_{\star}\gtrsim0.7\,M_{\odot}$.  Whereas, for targets with $\log(L_{\star}/L_{\odot})<-1$ and in the range $4.5 < M_{K_S} < 7.1$, we computed masses from the empirical magnitude--mass relationship from \citet{Mann2019}.

The masses of pre-main sequence stars were derived using stellar evolutionary tracks \citep[e.g.,][]{fang2021}. For this work, we converted the IDL code developed by \citet{pascucci2016} into Python. This code uses a Bayesian inference approach to estimate stellar mass, age, and associated uncertainties from the stellar \teff, $L_\text{bol}$, and a set of isochrones. The conditional likelihood function assumes uniform priors on the model properties and we propagated uncertainties on \teff and $L_\text{bol}$ from our photometric derivations. We used the non-magnetic \citet{Feiden2016} tracks to ages as old as 500\,Myr and included new tracks for magnetically-active stars. These new tracks are more appropriate for young M dwarfs ($M_{\star} \lesssim 0.5$\,\Msun) as shown, for example, by \citet{Simon2019} who compared masses from evolutionary tracks to dynamical masses and found that non-magnetic tracks systematically underestimate the pre-main sequence M dwarf masses by $\sim$50\%.

\section{Stellar Catalog Validation}\label{sec:compare}
Our homogeneous catalog of photometrically-derived \teff, $L_{\star}$, $R_{\star}$, and $M_{\star}$ for pre-main sequence and main sequence stars can be found in Tables~\ref{tab:pms} and \ref{tab:ms}. In the following sections, we performed a few validation checks of our derived stellar properties to those derived by other methods.

\begin{deluxetable*}{ccrcrccc}

\tablecaption{Pre-main sequence stellar properties.\label{tab:pms}}

\tablehead{\colhead{TIC} & \colhead{Cluster} & \colhead{Distance} & \colhead{$T_{\mathrm{eff}}$} & \colhead{log$L_{\star}$} & \colhead{$R_{\star}$} & \colhead{$M_{\star}$ (non-mag)} & \colhead{$M_{\star}$ (mag)}\vspace{-6pt} \\ 
\colhead{} & \colhead{} & \colhead{(pc)} & \colhead{(K)} & \colhead{($L_{\odot}$)} & \colhead{($R_{\odot}$)} & \colhead{($M_{\odot}$)} & \colhead{($M_{\odot}$)} } 

\startdata
138901588 & 32Ori & $100.097_{-0.244}^{+0.210}$ & $3209\pm42$ & $-1.414\pm0.020$ & $0.521\pm0.018$ & $0.174_{-0.174}^{+0.280}$ & $0.302_{-0.302}^{+0.320}$ \\
302417238 & 32Ori & $91.507_{-0.229}^{+0.271}$ & $3277\pm44$ & $-1.673\pm0.018$ & $0.422\pm0.014$ & $0.204_{-0.204}^{+0.296}$ & $0.372_{-0.372}^{+0.325}$ \\
365747593 & 32Ori & $98.177_{-0.126}^{+0.153}$ & $5743\pm79$ & $0.227\pm0.018$ & $1.311\pm0.044$ & $1.202_{-0.194}^{+0.277}$ & $1.122_{-0.026}^{+0.181}$ \\
449260853 & 32Ori & $91.516_{-0.154}^{+0.155}$ & $3183\pm42$ & $-1.298\pm0.019$ & $0.578\pm0.019$ & $0.162_{-0.162}^{+0.276}$ & $0.275_{-0.275}^{+0.317}$ \\
455029978 & 32Ori & $27.625_{-0.019}^{+0.018}$ & $3221\pm43$ & $-1.541\pm0.019$ & $0.465\pm0.015$ & $0.174_{-0.174}^{+0.284}$ & $0.309_{-0.309}^{+0.327}$ \\
19699155 & 118Tau & $108.085_{-0.207}^{+0.226}$ & $5379\pm74$ & $0.559\pm0.019$ & $2.192\pm0.077$ & $1.862_{-1.115}^{+1.201}$ & $1.549_{-0.606}^{+0.713}$ \\
54006139 & 118Tau & $87.904_{-0.131}^{+0.139}$ & $4594\pm62$ & $-0.277\pm0.018$ & $1.149\pm0.040$ & $1.047_{-0.024}^{+0.072}$ & $0.955_{-0.066}^{+0.022}$ \\
54185108 & 118Tau & $106.563_{-0.189}^{+0.201}$ & $3216\pm43$ & $-0.875\pm0.812$ & $0.620\pm0.020$ & $0.195_{-0.195}^{+0.292}$ & $0.302_{-0.302}^{+0.313}$ \\
62632828 & 118Tau & $95.540_{-0.357}^{+0.289}$ & $3063\pm52$ & $-1.727\pm0.020$ & $0.410\pm0.014$ & $0.105_{-0.105}^{+0.227}$ & $0.191_{-0.191}^{+0.285}$ \\
1364042 & ABDMG & $44.350_{-0.033}^{+0.029}$ & $3273\pm44$ & $-1.632\pm0.019$ & $0.434\pm0.014$ & $0.204_{-0.204}^{+0.296}$ & $0.372_{-0.372}^{+0.317}$ \\
\enddata


\tablecomments{Table~\ref{tab:pms} is published in its entirety in machine-readable format with additional columns. A small portion is shown here for guidance regarding its form and content.}


\end{deluxetable*}

\begin{deluxetable*}{ccrcrcc}

\tablecaption{Main sequence stellar properties.\label{tab:ms}}

\tablehead{\colhead{TIC} & \colhead{Cluster} & \colhead{Distance} & \colhead{$T_{\mathrm{eff}}$} & \colhead{log$L_{\star}$} & \colhead{$R_{\star}$} & \colhead{$M_{\star}$}\vspace{-6pt}  \\ 
\colhead{} & \colhead{} & \colhead{(pc)} & \colhead{(K)} & \colhead{($L_{\odot}$)} & \colhead{($R_{\odot}$)} & \colhead{($M_{\odot}$)} } 

\startdata
4069456 & 32Ori & $41.708_{-0.024}^{+0.026}$ & $5290\pm141$ & $-0.317\pm0.013$ & $0.824\pm0.047$ & $0.823\pm0.065$ \\
11085881 & 32Ori & $172.279_{-1.616}^{+1.746}$ & $3454\pm92$ & $-1.796\pm0.017$ & $0.363\pm0.013$ & $0.352\pm0.012$ \\
147799311 & 32Ori & $111.591_{-0.134}^{+0.134}$ & $5932\pm158$ & $0.239\pm0.015$ & $1.248\pm0.067$ & $1.119\pm0.088$ \\
284864375 & 32Ori & $58.250_{-0.051}^{+0.055}$ & $3836\pm102$ & $-1.117\pm0.016$ & $0.607\pm0.020$ & $0.598\pm0.019$ \\
371691843 & 32Ori & $36.780_{-0.017}^{+0.022}$ & $3411\pm91$ & $-1.888\pm0.015$ & $0.334\pm0.011$ & $0.314\pm0.010$ \\
408042385 & 32Ori & $44.120_{-0.047}^{+0.046}$ & $6273\pm168$ & $0.234\pm0.015$ & $1.109\pm0.062$ & $1.118\pm0.088$ \\
433143783 & 32Ori & $49.707_{-0.058}^{+0.060}$ & $4809\pm129$ & $-0.655\pm0.015$ & $0.680\pm0.038$ & $0.683\pm0.054$ \\
443750439 & 32Ori & $55.825_{-0.075}^{+0.078}$ & $3253\pm87$ & $-2.012\pm0.014$ & $0.308\pm0.010$ & $0.286\pm0.009$ \\
6749695 & 118Tau & $54.639_{-0.044}^{+0.057}$ & $3686\pm98$ & $-1.228\pm0.015$ & $0.572\pm0.018$ & $0.568\pm0.018$ \\
60511067 & 118Tau & $51.610_{-0.047}^{+0.042}$ & $3312\pm88$ & $-1.791\pm0.016$ & $0.370\pm0.013$ & $0.358\pm0.011$ \\
\enddata


\tablecomments{Table~\ref{tab:ms} is published in its entirety in machine-readable format with additional columns. A small portion is shown here for guidance regarding its form and content.}


\end{deluxetable*}

\subsection{Comparison Between Photometric and Spectroscopic Stellar properties}
To test the reliability of our photometrically-derived properties, we compared them to values from the GALAH DR3 \citep{Buder2021}, APOGEE DR17 \citep{Abdurro'uf2022apogeedr17}, and LAMOST DR8\footnote{\url{http://www.lamost.org/dr8/}} \citep{Cui2012} survey catalogs. For each survey, we imposed some simple quality cuts. For GALAH, we used the recommended quality cuts\footnote{\url{https://www.galah-survey.org/dr3/using_the_data/}} of SNR $>$ 30 in channel 3, and stellar parameter and iron abundance flags equal to zero, indicating no known problems with derived stellar properties. For APOGEE, we imposed a cut requiring no flags on stellar parameters (\texttt{STARFLAG=0}). For LAMOST, we imposed a SNR cut in $g$-band for AFGK stars and in $i$-band for M stars between 20 and 999, and made sure errors for \teff, $\log g$, and [Fe/H] or [M/H] were not -9999, the default value for poor quality measurements.

\begin{figure*}[!htb]
\gridline{\fig{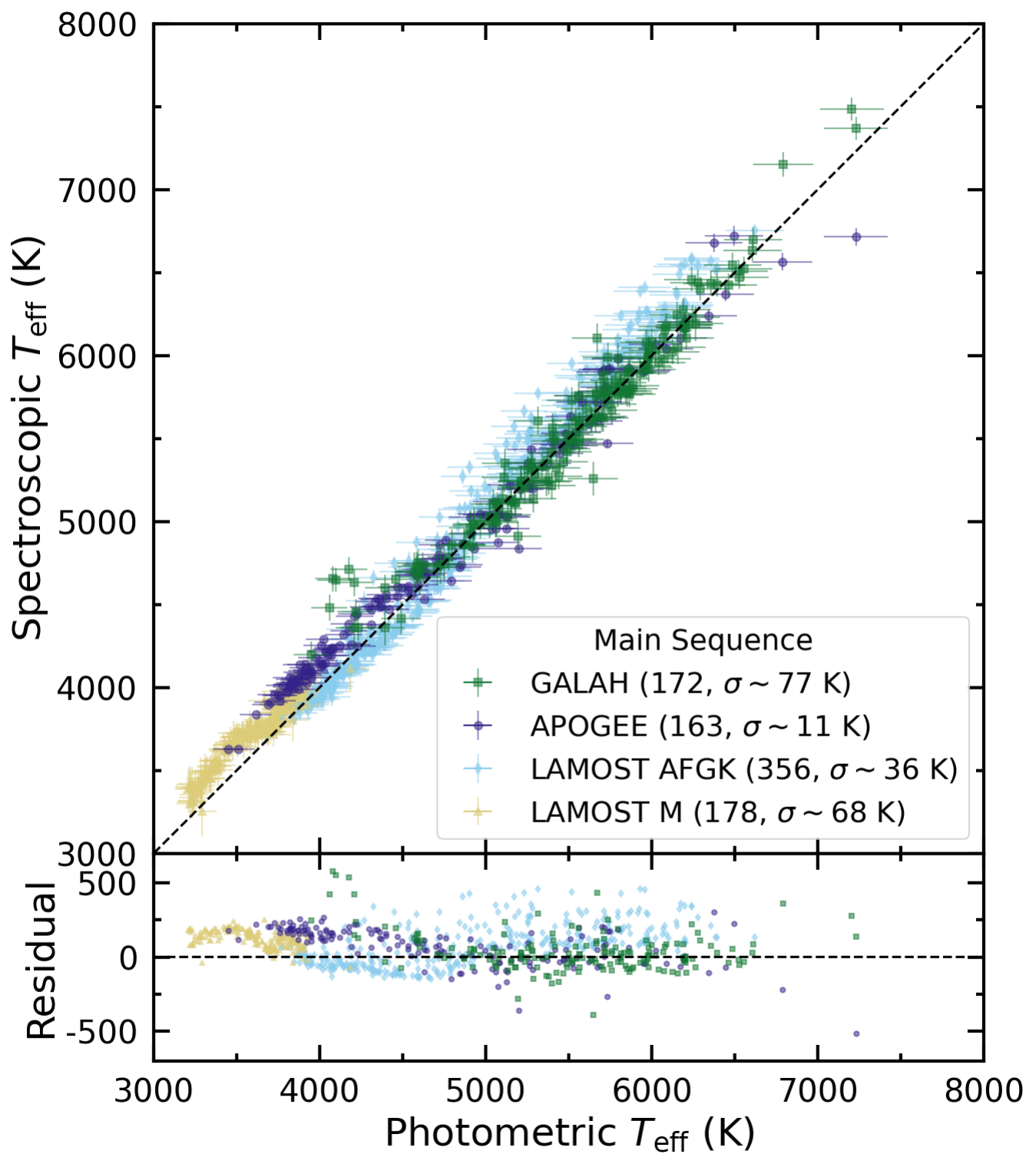}{0.48\textwidth}{}
          \fig{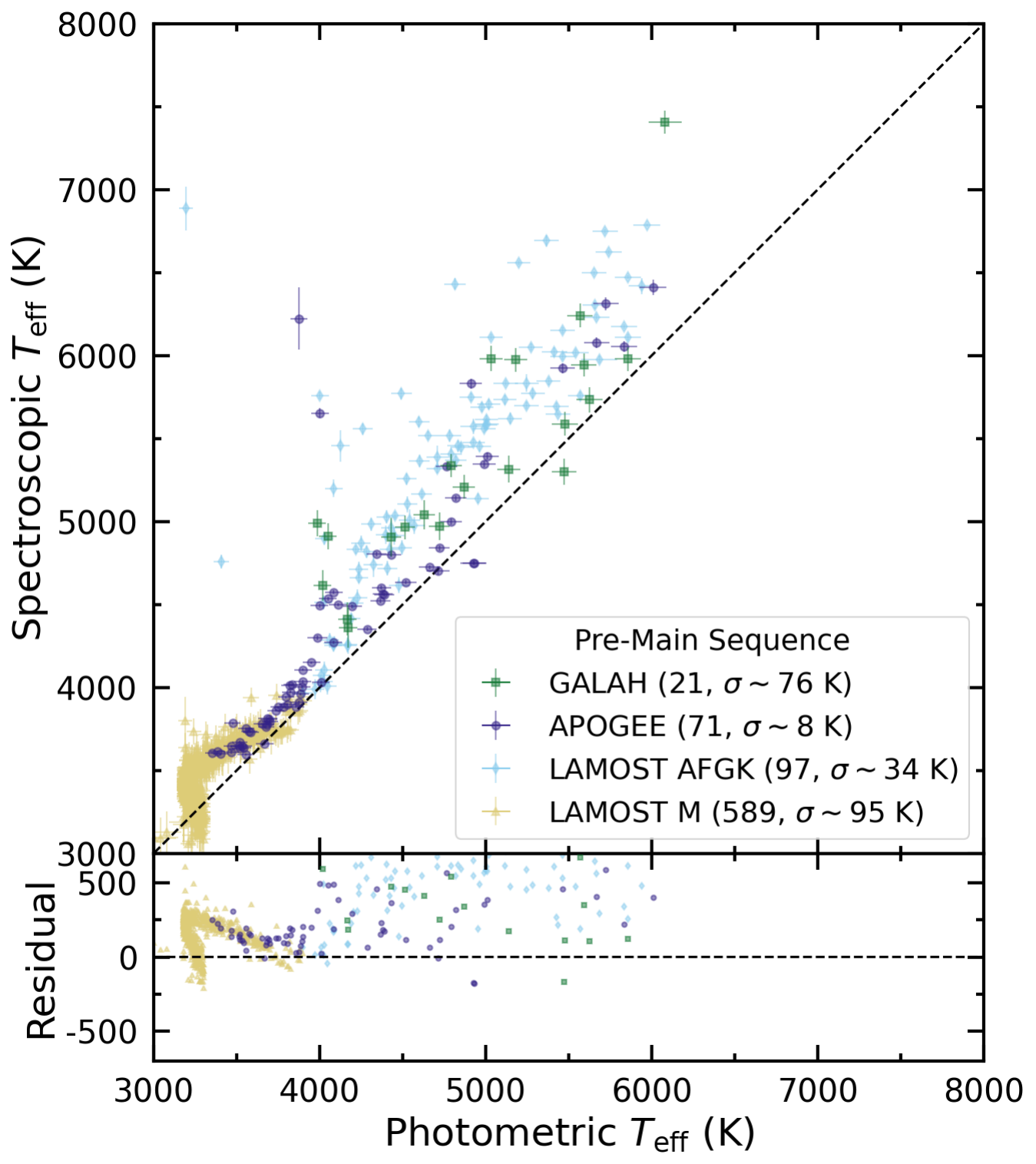}{0.48\textwidth}{}}
\vspace{-25pt}
\caption{Comparison between spectroscopically derived \teff from GALAH, APOGEE, and LAMOST to our photometrically derived \teff values for both main sequence fitting (left) and pre-main sequence fitting (right). For each survey, we list the number of targets and the median spectroscopic \teff\ uncertainty in parentheses.}\label{fig:teffcomp}
\end{figure*}

Each spectroscopic catalog contains \gaia DR3 IDs, so we cross-matched our TESS targets to each catalog based on this ID. For our main sequence sample, this yielded an overlap of 172 GALAH targets, 163 APOGEE targets, 356 LAMOST AFGK targets, and 178 LAMOST M targets. For our pre-main sequence sample, this yielded an overlap of 21 GALAH targets, 71 APOGEE targets, 97 LAMOST AFGK targets, and 589 LAMOST M targets. As can be seen in Figure~\ref{fig:teffcomp}, there is agreement between spectroscopically and photometrically derived temperatures for main sequence stars above $\sim$4000\,K, where the standard deviation of the difference in \teff is 138\,K, about the same as our 149\,K median \teff uncertainty for these stars. Photometric \teff values for main sequence stars below $\sim$4000\,K are on average 118\,K lower than the spectroscopic measurements, which is slightly higher than our median \teff uncertainty of 93\,K for these cool stars. \citet{Andrae2018} and \citet{Dressing2019} identified a similar trend and suggested either extinction or strong molecular features in M dwarfs causing the discrepancy. Since we do not see a similar \teff offset for warmer stars, we assume the most likely cause is due to the strong molecular features in M dwarfs which can make it difficult to fit precise stellar parameters, rather than extinction. Pre-main sequence stars typically have higher spectroscopic \teff values than photometric values. The likely cause of this discrepancy is the use of main-sequence stellar models in the spectroscopic parameter fitting which are not necessarily be appropriate for most young stars.



\begin{figure*}[!htb]
\gridline{\fig{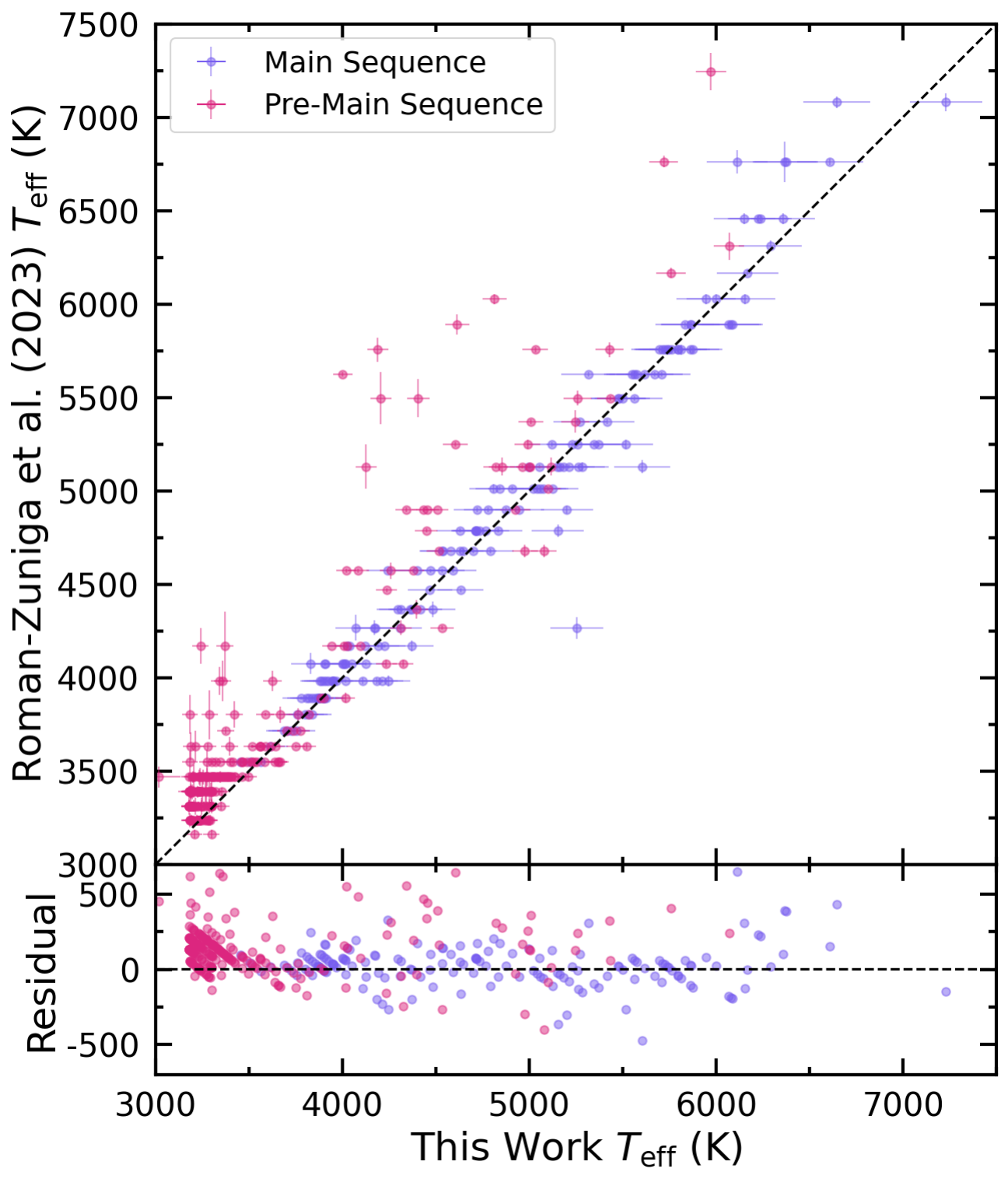}{0.48\textwidth}{}
          \fig{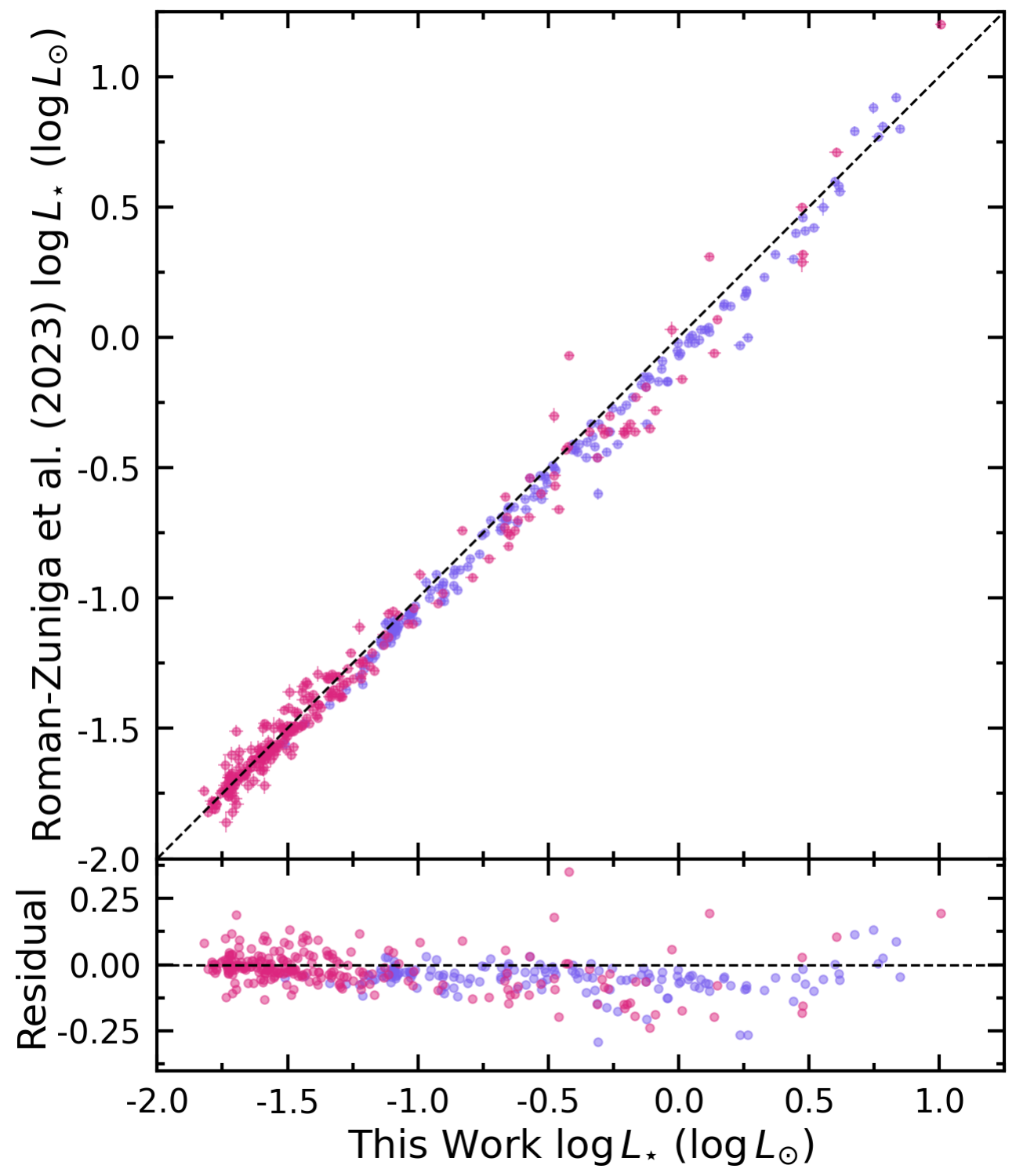}{0.48\textwidth}{}}
\vspace{-25pt}
\gridline{\fig{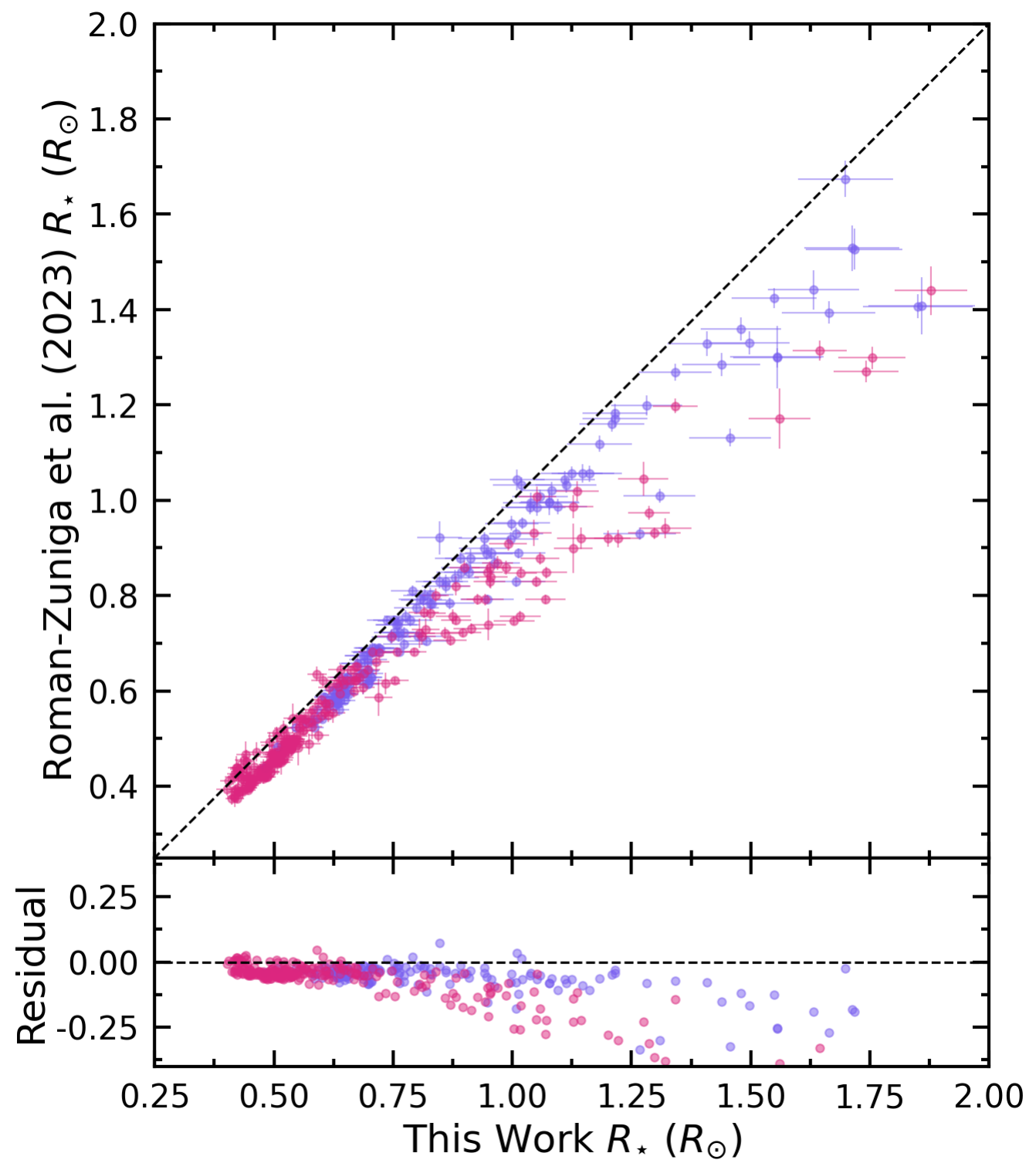}{0.48\textwidth}{}
          \fig{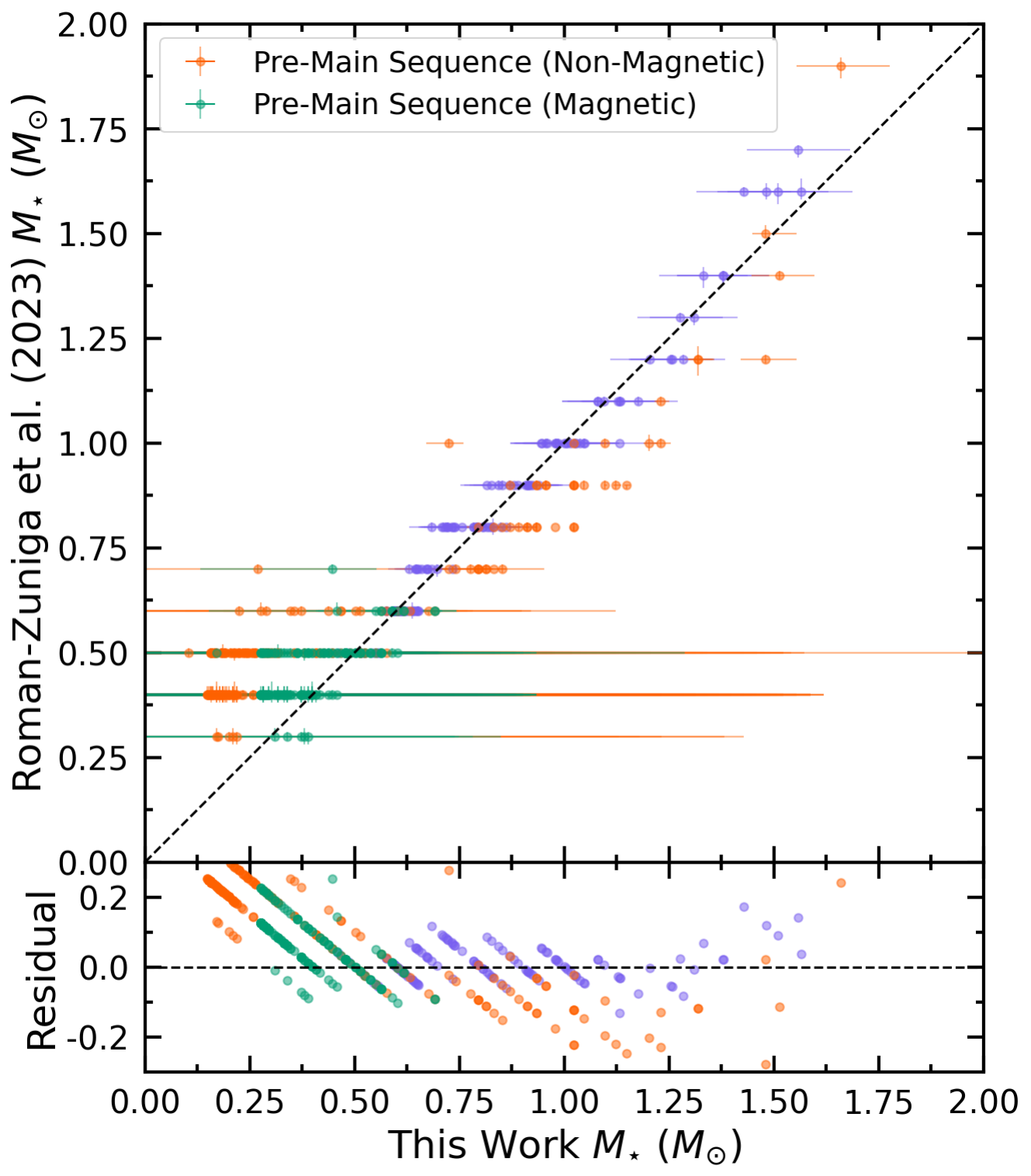}{0.48\textwidth}{}}
\vspace{-25pt}
\caption{Comparison between spectroscopically derived \teff, luminosity, stellar radius, and mass derived from APOGEE spectra as in \citet{RomanZuniga2023} and our photometrically-derived \teff values for both main sequence and pre-main sequence stars.}\label{fig:apocomp}
\end{figure*}

We also compared our stellar properties with those derived by \citet{RomanZuniga2023} using spectra from APOGEE DR16 and 17. \citet{RomanZuniga2023} identified a sample of 3,360 young stars for which they derived \teff, $\log g$, [Fe/H], $L_{\star}$, $M_{\star}$, and age using tools separate from the standard APOGEE pipeline. There are 168 main sequence, and 251 pre-main sequence stars that overlap with our sample. As illustrated in Figure~\ref{fig:apocomp}, we found that our photometrically-derived \teff values are consistent with those derived from APOGEE spectra. However, for stars brighter than $\log L_{\odot} \approx -1$, \citet{RomanZuniga2023} derived increasingly larger luminosities; the effect of higher luminosities can be seen propagated into the derivation of higher stellar radii. We attribute this disagreement in the luminosity and radii of earlier-type stars to the specific manner in which extinction is taken into account in the two works since the difference in \teff ($\sim$150\,K) between the two works is not significant enough to make a difference in the derivation of the radii, and both works use \gaia DR3 magnitudes. While our work relies on more recent dust maps from \citet{Green2019} and \citet{Schlafly2011}, \citet{RomanZuniga2023} took a more empirical approach and did not use dust maps. They estimated the visual extinction for each source and established a confidence range through a Monte Carlo method. This was achieved by minimizing the differences between the extinction-corrected colors and the expected intrinsic colors from \citet{Luhman2020}. To correct the observed colors, they applied the extinction law of \citet{Cardelli1989}, assuming a canonical interstellar reddening law with Rv = 3.1 for all regions. Their luminosities were derived using the extinction-corrected J magnitude, the bolometric correction for PMS stars from \citetalias{pecautmamajek2013}, and Gaia EDR3 geometric distance estimations from \citet{bailerjones2021}. \citet{RomanZuniga2023} derived their masses via Monte-Carlo sampling and interpolation within the \texttt{PARSEC-COLIBRI} evolutionary model grid \citep{Bressan2012}. We did, however, find that main sequence mass measurements are consistent with a median offset of 0.004\,\Msun, well below our typical measurement uncertainty of 0.058\,\Msun. For pre-main sequence stars, our masses derived using magnetic isochrones are visually more consistent with those derived by \citet{RomanZuniga2023}, however, for stars in the range of M dwarf masses ($\lesssim$0.6\,\Msun), the isochronal mass uncertainties are very large (See Figure~\ref{fig:apocomp}).

Our comparisons with spectroscopically-derived properties show reasonable agreement, with some deviations which we attribute to different methodology such as how we accounted for extinction or stellar model grids. Our measurements can be improved in the future with a large uniform spectroscopic survey of thousands of young stars with a wide range of ages and across all spectral types.

\subsection{Main Sequence vs. Pre-Main Sequence Stellar properties}
\begin{figure*}[!htb]
    \centering
    \includegraphics[width=1.0\linewidth]{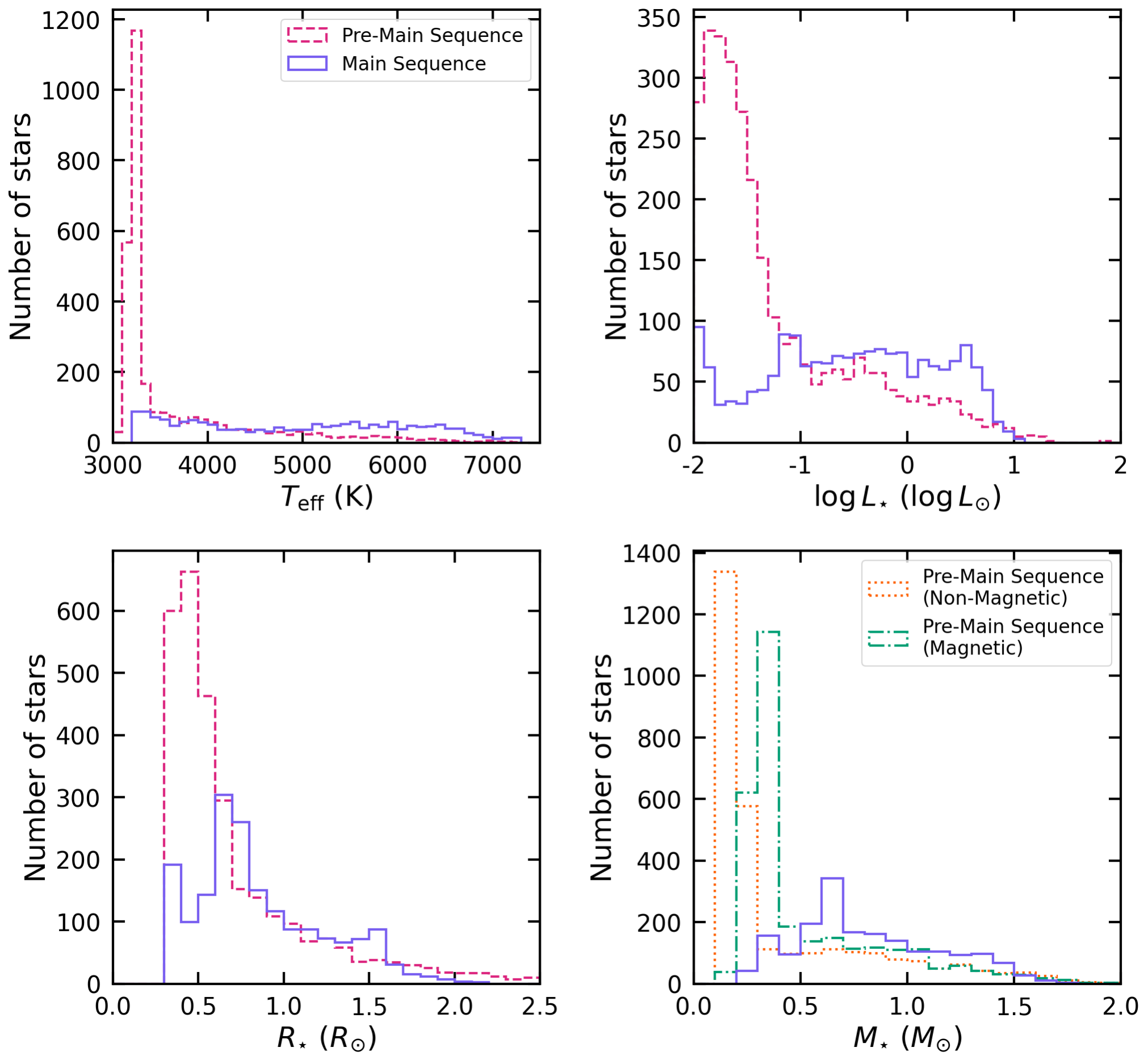}
    \caption{Histograms comparing the stellar properties of our pre-main sequence and main-sequence stars showing effective temperatures (upper left), luminosities (upper right), radii (lower left), and masses (lower right). In all panels, the main sequence distribution is depicted in purple. For the effective temperatures, luminosities, and radii panels, the pre-main sequence distribution is in pink. In the mass panel, we show the pre-main sequence masses from non-magnetic (orange) and magnetic (green) stellar isochrones to highlight the different distributions.
    }\label{fig:starhist}
\end{figure*}

We also compared the stellar properties between our pre-main sequence and main sequence populations (see Figure~\ref{fig:starhist}). It is important to note that the median age of the stars in our sample is $\sim$45--50\,Myr, at which point a majority of the M dwarfs have not yet reached the main sequence. As such, our sample has a much larger number of pre-main sequence M dwarfs than main sequence M dwarfs. Overall, we see a significant decrease in the fraction of pre-main sequence earlier-type stars. Comparing stellar masses, specifically in the 0.1--0.5\,\Msun\ region, we found that masses derived using non-magnetic stellar isochrones were lower by $\sim$50\% compared to those derived using magnetic stellar isochrones. This is consistent with \citet{Simon2019}, where they found that non-magnetic stellar isochrones do not properly account for the strong magnetic processes dominant in lower mass stars. It is also important to note that for stars with  $M_{\star}>0.5$\,\Msun, the pre-main sequence masses derived from non-magnetic stellar isochrones match those derived using non-magnetic stellar isochrones and follow the same overall distribution as the main sequence masses.

\section{The Effect of Stellar properties on Survey Completeness}\label{sec:completeness}

The intrinsic occurrence rate of planets ($\eta$) can be calculated from the fraction of stars with detected planets in a survey and the survey completeness as follows:
\begin{equation}
\eta = \frac{{n_\text{p}}}{n_\star} \times \frac{1}{comp_\text{bin}}
\end{equation}
where $comp_\text{bin}$ is the survey completeness evaluated in a discrete radius and orbital period bin, $n_p$ is the number of detected planets in the bin and $n_\star$ is the number of surveyed stars. The survey completeness is computed by combining the detection efficiency (calculated using injection-recovery tests), and the geometric transit probability, which is given by
\begin{equation}
f_\text{geo} = \frac{R_\star}{a}
\end{equation}
where $R_\star$ is the stellar radius, and $a$ is the average semi-major axis, which is calculated from the orbital period using Kepler's third law. The uncertainty on the occurrence rate was calculated from the square root of the number of detected planets in the bin, assuming Poisson statistics.

In our preliminary analysis of short-period planets in young clusters \citep{fernandes2022}, we searched five clusters with ten known transiting planets: Tucana-Horologium Association, IC~2602, Upper Centaurus Lupus, Ursa Major, and Pisces-Eridanus. We ran the TESS Primary Mission Full Frame Images (FFIs; 30-min cadence) through our pipeline \pterodactyls \citep{rachel_fernandes_2022_6667960} and recovered seven of the eight confirmed planets and one of the two planet candidates. Here, it is important to reiterate that these clusters were solely selected to be used as a test sample to evaluate the effectiveness of our code in recovering these known planets. We specifically focused on sub-Neptunes and Neptunes (0.017--0.055 \rprs\ or 1.8--6\,\Rearth, assuming a solar radius for the host star) with orbital periods $<12.5$\,days (about half a TESS sector) to better understand the primordial population of sub-Neptunes and Neptunes before they are stripped of their atmospheres. Using \gaia, we took into account the flux contamination prominent in young cluster environments, and computed our detection efficiency in \rprs\ space because, at the time, most stars in our clusters lacked stellar properties. Given the lack of a homogeneous stellar catalog, we previously included stars of all spectral types in the analysis of our detection efficiency. With an average detection efficiency of 9\% and geometric transit probability ($f_\text{geo}$) of 0.1 (at a geometric mean orbital period of 3.5\,days and assuming a solar-type star), we computed an occurrence rate of 49$\pm$20\% for sub-Neptunes and Neptunes in our biased sample of young clusters. This is much higher than the \kepler Gyr-old FGK (Sun-like) occurrence rate of 6.8$\pm$0.3\% in the same planet radius and orbital period bin. In this section, we revisited that number using the radii and masses calculated from our homogeneous approach.

\begin{figure*}[!htb]
\begin{center}
\minipage{0.5\textwidth}
  \includegraphics[width=\linewidth]{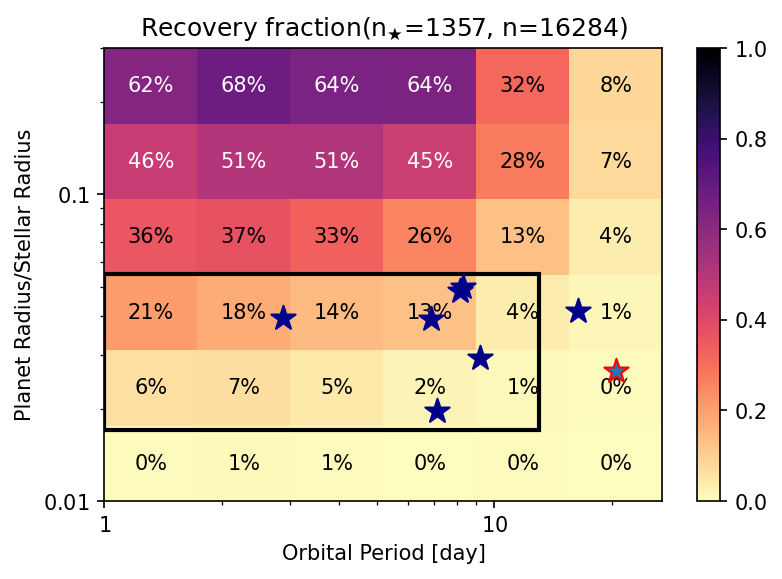}
\endminipage\hfill
\minipage{0.5\textwidth}
  \includegraphics[width=\linewidth]{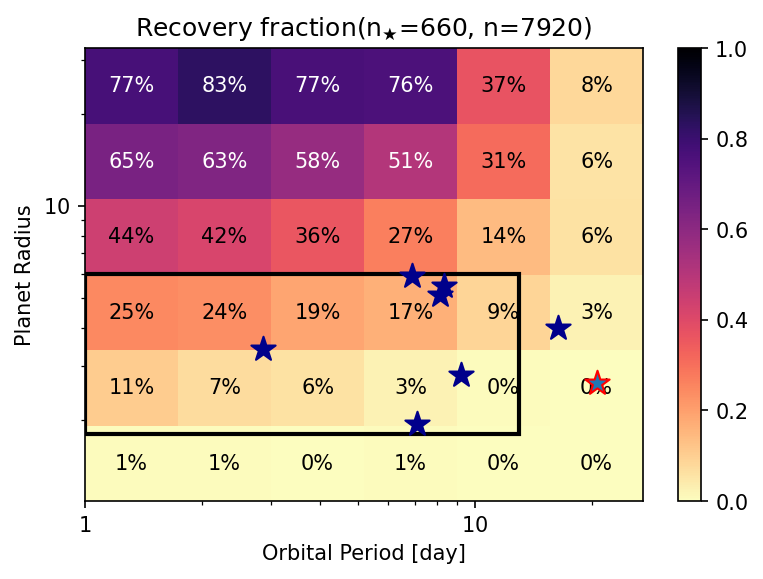}
\endminipage\hfill
\minipage{0.5\textwidth}
  \includegraphics[width=\linewidth]{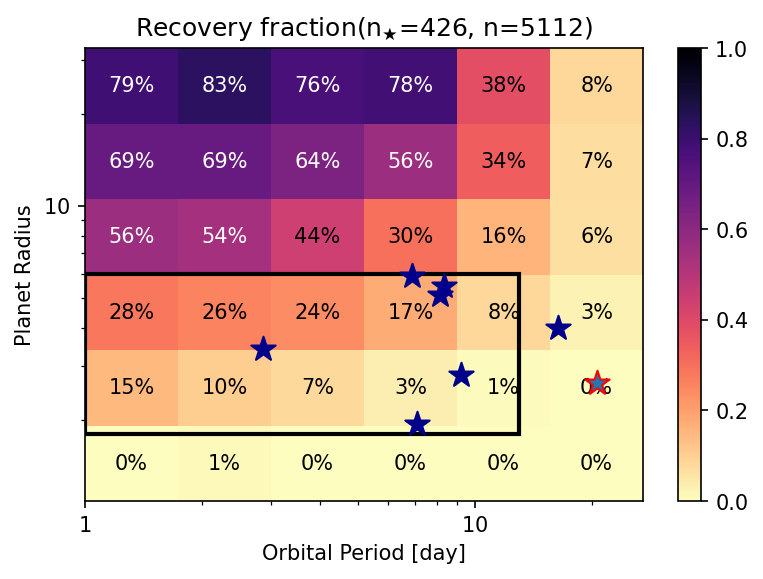}
\endminipage
\end{center}
\caption{Comparison of detection efficiencies of the sample analyzed in \citet{fernandes2022}. The confirmed planets are depicted using blue stars, except for TOI~1726\,b whose recovery required changing \pterodactyls. Darker bins represent regions of higher detection efficiency. ${n_\star}$ is the total number of stars in our sample while \textit{n} is the total number of injections done. Black box denotes the bin over which the intrinsic occurrence rates were calculated i.e., sub-Neptunes and Neptunes (0.017--0.055 \rprs\ or 1.8--6\,\Rearth, assuming a solar radius) within an orbital period of 12.5\,days (about half a TESS sector). \textbf{Top Left:} Detection efficiency of all 1357 stars in \rprs-period space before stellar characterization. \textbf{Top Right:} Detection efficiency in planet radius-orbital period space of 660 stars for which we were able to compute stellar properties. \textbf{Bottom:} Detection efficiency in planet radius-orbital period space of 426 FGK (0.55--1.63\,\Msun) stars.}
\label{fig:deteff_comp}
\end{figure*}

In this paper, we computed stellar properties for 660 out of 1357 stars studied in \citet{fernandes2022}, with the remaining stars being too faint or lacking necessary 2MASS and \gaia DR3 photometry. After revising our sub-Neptune and Neptune regime detection efficiency with these updated stellar properties, we found a slight increase in the average detection efficiency from 9\% to 10\% (Figure~\ref{fig:deteff_comp}). However, since the total number of stars decreased by 50\%, the occurrence rate of sub-Neptunes and Neptunes increased to 90$\pm$37\%. For a better comparison with \kepler's Gyr-old planet population orbiting Sun-like stars, we analyzed only young stars of FGK spectral type (426 stars; 0.55--1.63\,\Msun). We observed an average detection efficiency of 15\% in the young sub-Neptune and Neptune regime, leading to an occurrence rate of 93$\pm$38\%. This is comparable to the 90$\pm$37\% for all spectral types, which may be partially due to the removal of faint M dwarf stars that make it harder to detect transiting planets with TESS \citep{Kunimoto2022}.

While this occurrence rate is indeed higher than that of \kepler's Gyr-old population, even when accounting for the fraction of evaporated sub-Neptunes \citep{bergsten2022}, it is still heavily biased given that we only looked at clusters that are known to have planets. As such, including clusters without known detected planets would lead to a lower occurrence rate. In fact, given that we now know that there are $\sim$2300 Sun-like stars, and 15 published planets in our sample of nearby ($<$200\,pc) young clusters and moving groups, assuming the same detection efficiency, the occurrence would drop to 43$\pm$11\%, which is effectively the same as the value of 49$\pm$20\% that we computed in \citet{fernandes2022}. This observed increase in the occurrence rate of sub-Neptunes and Neptunes could indicate a surplus of these planets at young ages. In future work, as we expand our sample to include data from all nearby clusters and moving groups, it will be important to determine whether this increase in occurrence rate persists when working with a non-biased sample. If it does, it could be due to the fact that the atmospheres of these young sub-Neptunes and Neptunes have not yet been stripped by photoevaporation or core-powered mass loss.


\section{Summary and Discussion}\label{sec:summary}
The discovery and characterization of transiting exoplanets is strongly dependent on our understanding of their host star properties. In this paper, we used \gaia and 2MASS photometry along with stellar (magnetic and non-magnetic) isochrones in order to derive a homogeneous set of stellar properties, specifically \teff, $L_{\star}$, $R_{\star}$, and $M_{\star}$ for 4,865 stars (1,824 main sequence, and 3,041 pre-main sequence) from 31 nearby ($\sim$200\,pc) young ($<$1\,Gyr) clusters and moving groups. We found that:

\begin{itemize}[leftmargin=*]
\item Our photometrically-derived \teff values generally agree within measurement uncertainties to those derived from spectroscopic surveys APOGEE and GALAH and LAMOST for both main sequence populations. There are some discrepancies, however, for pre-main sequence stars, which we attribute to spectroscopic surveys using stellar models tuned to main sequence stars.

\item Our \teff values typically agree within measurement uncertainties to those derived in \citet{RomanZuniga2023} using APOGEE spectra. Our non-M dwarf luminosities and radii increasingly deviate toward earlier-type stars, which we attribute to differences in accounting for interstellar extinction.

\item For FGK type stars, we found that there was no difference in the masses derived from non-magnetic vs. magnetic stellar isochrones. On the other hand, for pre-main sequence M dwarfs (0.1--0.5\,\Msun), our masses derived from magnetic stellar isochrones are visually more consistent with those derived in \citet{RomanZuniga2023}, however, our measurement uncertainties are on the order of 100\% for this population, making it difficult to conclusively rule out non-magnetic measurements. Though, \citet{Simon2019} found that non-magnetic stellar isochrones tend to underestimate stellar masses, which is consistent with what we see here.

\item Given that the median age of our sample is $\sim$45--50\,Myr, most of the M dwarfs in our sample have not yet reached the main sequence. We also note a marked decrease in the number of earlier-type pre-main sequence stars with respect to main sequence stars.

\item When we took stellar properties into account, there is an overall increase in our detection efficiency. This effect is particularly noticeable among sub-Neptunes/Neptunes (1.8--6\,\Rearth) where we found a $1.5\times$ increase in the detection efficiency, implying that better characterization of host star properties can lead to the recovery of more smaller transiting planets.
\end{itemize}

The occurrence rates of sub-Neptunes/Neptunes (1.8--6\,\Rearth) with orbital periods less than 12.5\,days for the 660 stars (out of 1357) for which we could compute stellar properties is 90$\pm$37\%. This is significantly higher than \kepler's Gyr-old occurrence rates of 6.8$\pm$0.3\% even when accounting for evaporated sub-Neptunes \citep{bergsten2022}, as well as our previously calculated 49$\pm$20\% \citep{fernandes2022} when considering stars of all spectral types in \rprs\ space. When considering only young stars of FGK spectral type (426 stars; 0.55--1.63\,\Msun), we computed an occurrence of 93$\pm$38\%, which is effectively the same as 90$\pm$37\% when considering all stars with stellar properties. While the number of detected planets has not changed, the difference can be attributed to two factors: (1) the total number of only FGK stars is $\sim$1.5 times lower, and (2) the detection efficiency is $\sim$1.5 times higher.
While the planet occurrence rate we calculated for our sample is higher than that for \kepler's Gyr old stars, we realize that this value is biased since we only considered clusters with confirmed/candidate planets. We can further improve upon this by including all of the $>$30 young short-period transiting planets as well as an unbiased sample of nearby clusters and moving groups using light curves from the TESS Extended Mission FFIs. Our homogeneously derived catalog of updated stellar properties will be a crucial input to the accurate estimation of the occurrence rates of young planets. With this young population of transiting, short-period planets, we hope to improve upon the intrinsic occurrence rates calculations, establish how the radius distribution of transiting exoplanets evolved over time, and therefore provide observational constraints on the mass loss mechanisms of planetary atmospheres.

\software{\pterodactyls \citep{rachel_fernandes_2022_6667960}, \texttt{NumPy} \citep{numpy}, \texttt{SciPy} \citep{scipy}, \texttt{Matplotlib} \citep{pyplot}, \eleanor{} \citep{feinstein2019eleanor}, \wotan{} \citep{hippke2019wotan}, \texttt{transitleastsquares} \citep{hippke2019optimized}, \texttt{vetting} \citep{hedges2021vetting}, \triceratops{} \citep{giacalone2020vetting}, \texttt{EDI-Vetter Unplugged} \citep{zink2019edivetter}, \exotic \citep{zellem2020utilizing}, \epos \citep{Mulders2018}}

\acknowledgements
R.B.F. and K.H-U. would like to thank the following individuals for their expertise, assistance and, invaluable insights throughout this work: F\'{a}bio Wanderley, David Ciardi, Travis Barman, and Tommi Koskinen.
I.P., G.B., and K. C. acknowledge support from the NASA Astrophysics Data Analysis Program under  grant  No.  80NSSC20K0446. G.D.M. acknowledges support from FONDECYT project 11221206, from ANID --- Millennium Science Initiative --- ICN12\_009, and the ANID BASAL project FB210003. Part of this research was carried out in part at the Jet Propulsion Laboratory, California Institute of Technology, under a contract with the National Aeronautics and Space Administration (80NM0018D0004). This paper includes data collected by the TESS mission. Funding for the TESS mission is provided by the NASA's Science Mission Directorate. This material is based upon work supported by the National Aeronautics and Space Administration under Agreement No. 80NSSC21K0593 for the program “Alien Earths”. The results reported herein benefited from collaborations and/or information exchange within NASA’s Nexus for Exoplanet System Science (NExSS) research coordination network sponsored by NASA’s Science Mission Directorate.
\clearpage

\bibliographystyle{apj}
\bibliography{main}
\clearpage
\appendix
\section{Summary of Young Clusters and Moving Groups}\label{sec:photometry}

\begin{table*}[!htb]
    \centering
    \begin{tabular}{|c|c|c|c|c|c|c|c|}
    \hline\hline
    Cluster/Moving Group & Distance (in pc) & Age (in Myr) & Total & \# in \gaia & \# in 2MASS & \# in PMS & \# in MS \\
    \hline\hline
118~Tau&100$\pm$10&$\sim$10&15&15&15&4&3\\
32~Ori&96$\pm$2&$22^{+4}_{-3}$&42&41&41&5&8\\
AB Doradus MG&$30^{+20}_{-10}$&$149^{+51}_{-19}$&596&594&539&79&113\\
Alpha Persei&$\sim$200&90&740&740&717&368&75\\
Argus&$\sim$120&40-50&38&38&37&5&14\\
Beta Pictoris&$30^{+20}_{-10}$&24$\pm$3&303&302&271&69&51\\
Blanco~1&253&132&489&489&481&98&189\\
Carina&60$\pm$20&$30^{+11}_{-7}$&123&123&114&46&12\\
CarinaNear&30$\pm$20&$\sim$200&183&183&161&20&27\\
Coma Bernices&$\sim$85&$562^{+98}_{-84}$&165&165&160&4&64\\
Columba&50$\pm$20&$42^{+6}_{-4}$&220&220&200&49&44\\
Eta Cha&95$\pm$1&11$\pm$3&18&18&18&7&0\\
Hyades&40-50&750$\pm$100&612&612&573&91&146\\
IC~2391&149$\pm$6&50$\pm$5&333&333&328&151&34\\
IC~2602&149$\pm$5&$46^{+6}_{-5}$&504&504&477&253&29\\
Lower Centaurus Crux&110$\pm$10&15$\pm$3&530&529&500&144&62\\
Mu Tau&150&60&566&566&509&143&70\\
NGC~2451&180-360&50-80&400&400&380&197&35\\
Octans&$130^{+30}_{-20}$&35$\pm$5&159&159&151&32&40\\
Pisces-Eridanus&80-226&120&254&254&250&48&114\\
Platais 8&130$\pm$10&60&35&35&33&4&12\\
Pleiades&134$\pm$9&112$\pm$5&1593&1590&1545&467&253\\
Praesepe&179$\pm$3&790$\pm$60&860&860&835&226&265\\
Tucana Horologium Association&$46^{+8}_{-6}$&45$\pm$4&212&211&205&85&15\\
TW Hydra&60$\pm$10&10$\pm$3&86&86&73&17&2\\
Upper Centaurus Lupus&130$\pm$20&16$\pm$2&936&934&903&248&101\\
Upper CrA&147$\pm$7&$\sim$10&41&41&38&21&1\\
Ursa Major&$\sim$25&414$\pm$23&17&17&16&0&5\\
Upper Scorpius&130$\pm$20&10$\pm$3&411&411&395&149&10\\
Volans Carina&75-100&$89^{+5}_{-7}$&86&86&82&8&25\\
XFOR&$\sim$100&$\sim$40&17&17&17&3&5\\
    \hline\hline
Total&&&10,585&10,573&10,064&3,041&1,824\\
    \hline\hline
    \end{tabular}
\caption{{Moving groups and clusters whose properties were derived in this work. Distance, age, and membership are from the
\citet{gagne2018banyan} and \citet{babusiaux2018gaia}.}}
\label{table:photometry}  
\end{table*}

\clearpage

\end{document}